\documentclass[11pt]{article}
\usepackage{graphicx}
\usepackage{url}

\newcommand{\BABARPubYear}    {06}
\newcommand{\BABARProcNumber} {049}
\newcommand{\SLACPubNumber} {11942}

\input pubboard/babarsym

\newcommand{\calB}{\ensuremath{\mathcal{B}}}
\newcommand{\timesix}{\ensuremath{\times10^{-6}}}
\newcommand{\rhoz}{\ensuremath{\rho^0}}
\newcommand{\rhop}{\ensuremath{\rho^+}}
\newcommand{\rhom}{\ensuremath{\rho^-}}
\newcommand{\etaprp}{\ensuremath{\eta^{(\prime)}}}
\newcommand{\fetapiz}{\ensuremath{\eta\piz}\xspace}
\newcommand{\etapiz}{\ensuremath{\Bz\ra\fetapiz}\xspace}
\newcommand{\Betapiz}{\ensuremath{\calB(\etapiz)}\xspace}
\newcommand{\fetapeta}{\ensuremath{\etapr\eta}}
\newcommand{\etapeta}{\ensuremath{\Bz\ra\fetapeta}}
\newcommand{\Betapeta}{\ensuremath{\calB(\etapeta)}}

\newcommand{\Betappiz}{\ensuremath{\calB(\Bz\ra\etapr \piz)}\xspace}

\long\def\inst#1{\par\nobreak\kern 4pt\nobreak
    {\it #1}\par\vskip 10pt plus 3pt minus 3pt}

\begin{document}
{\pagestyle{empty}

\begin{flushright}
SLAC-PUB-\SLACPubNumber \\
\babar-PROC-\BABARPubYear/\BABARProcNumber \\
July 2006 \\
\end{flushright}

\par\vskip 4cm

\begin{center}
\Large \bf \boldmath
Probing QCD with rare charmless $B$ decays
\end{center}
\bigskip

\begin{center}
\large 
W. Gradl\\
School of Physics \\
The University of Edinburgh, Edinburgh EH9 3JZ, UK\\
(from the \babar{} Collaboration)
\end{center}
\bigskip \bigskip

\begin{center}
\large \bf Abstract
\end{center}
  Rare charmless hadronic $B$ decays are a good testing
  ground for QCD.  In this paper we describe a selection of new
  measurements made by the \babar{} and BELLE collaborations.

\vfill
\begin{center}
Contributed to the Proceedings of the \\
$14^\mathrm{th}$ International Workshop On Deep Inelastic Scattering (DIS 2006), \\
20-24 Apr 2006, Tsukuba, Japan
\end{center}

\vspace{1.0cm}
\begin{center}
{\em Stanford Linear Accelerator Center, Stanford University, 
Stanford, CA 94309} \\ \vspace{0.1cm}\hrule\vspace{0.1cm}
Work supported in part by Department of Energy contract DE-AC03-76SF00515.
\end{center}

\section{Introduction}
Rare charmless hadronic $B$ decays are a good testing ground for the
standard model.  
The dominant amplitudes contributing to this class of $B$ decays
are CKM suppressed tree diagrams and $b\to s$ or $b\to d$ loop
diagrams (`penguins').  These decays can be used to study interfering
standard model (SM) amplitudes and CP violation.  They are sensitive
to the presence of new particles in the loops, and they provide
valuable information to constrain theoretical models of $B$ decays.

The $B$ factories \babar{} at SLAC and Belle at KEK produce $B$
mesons in the reaction $\epem \to \FourS \to \BB$.  So far they have
collected integrated luminosities of about $600\invfb$ and
$380\invfb$, respectively.  The results presented here are based on
subsets of about $200$--$350 \invfb$ and are preliminary unless a
journal reference is given.

\section{$\Delta S$ from rare decays}

The time-dependent CP asymmetry in $B$ decays is observed as an
asymmetry between \Bz{} and \Bzb{} decay rates into CP eigenstates $f$
\begin{equation}
  \label{eq:cpv}
          \mathcal{A}_{cp} (\Delta t) =  \frac{\Gamma(\Bzb\to f) -
            \Gamma(\Bz \to f)}{\Gamma(\Bzb\to f) + \Gamma(\Bz \to f)}
          = S_f \sin\Delta m_d \Delta t - C_f \cos\Delta m_d \Delta t
          , \nonumber
\end{equation}
where $\Delta m_d = 0.502\pm0.007 \mathrm{ps^{-1}}$ and $\Delta t$ is
the time difference between the decays of the two neutral $B$ mesons
in the event.  The coefficients $S_f$ and $C_f$ depend on the final
state $f$; for the `golden' decay $\Bz\to\jpsi\KS$, for example,
$S_{\jpsi\KS} = \sin2\beta$, $C_{\jpsi\KS} = 0$.  Here, $\beta \equiv
\phi_1$ is one of the angles of the unitarity triangle of the CKM
matrix.  In general, the presence of more than one contributing
amplitude for the decay can introduce additional phases, such that $S_f$
measured in such a decay deviates from the simple $\sin2\beta$.  There
are intriguing hints in experimental data that $S_f$ is smaller than
$\sin 2\beta$ in $B$ decays involving the transition $b\to\qqbar s$,
like $\Bz\to\phi\Kz$, $\Bz\to\etapr\Kz$, or $\Bz\to\piz\Kz$.  However,
for each of these final states the SM contribution to $\Delta
S_f\equiv S_f - \sin2\beta$ from sub-dominant amplitudes needs to be
determined in order to draw a conclusion about the presence of any new
physics.  Typically, models prefer $\Delta S_f > 0$
\cite{Beneke:2005pu,Cheng:2005bg}, while for the final state
$\etapr\KS$, a small, negative $\Delta S_f$ is
expected\cite{Williamson:2006hb}.  Measuring $B$ decays which are 
related to the ones above by approximate SU(3) flavor or isospin
symmetries helps to constrain the expected $\Delta S_f$.

The sub-dominant contributions to $\Bz\to\phi\Kz$ can be constrained
using SU(3) flavor relations\cite{Grossman:2003qp}.  This requires
branching fraction measurements for eleven decay channels
($\Kstarz\Kzb, \Kstarzb\Kz$, and $hh^\prime$ with $h = \rhoz,
\omega, \phi$ and $h^\prime = \piz, \eta, \etapr$).
\babar{} has measured an upper limit\cite{Aubert:2006wu} for the sum
$\calB(\Kstarz\Kzb) + \calB(\Kstarzb\Kz) < 1.9\timesix$
and an updated upper limit\cite{Aubert:2006nn} for $\phi\piz$ of
$\calB(\phi\piz) < 2.8\times 10^{-7}$.  This allows one to place a
bound on $|\Delta S_{\phi\Kz}| < 0.43$.

The decays $\Bz\to\etaprp\piz, \etapr\eta$ can be used to constrain
the SM pollution in \Betaprkz, The expected branching fractions are
between $0.2$ and $1 \times 10^{-6}$ for \etaprp\piz{} and $0.3$ -
$2\times 10^{-6}$ for $\etapr\eta$. Using 211\invfb{} of data,
\babar{} sets the following upper limits\cite{Aubert:2006qd} at 90\%
confidence level (C.L.)  in units of $10^{-6}$: $\Betapiz < 1.3$,
$\Betapeta < 1.7$, $\Betappiz <2.1$, while Belle\cite{Schumann:2006bg}
measures $\Betappiz = (2.79 ^{+ 1.02 + 0.25} _ {-0.96 -0.34})\timesix$
with $386\times 10^6$ analyzed $\BB$ pairs.  Following Ref.
\cite{Gronau:2004hp}, the expected improvement on the prediction of
$\Delta S_{\etapr\KS}$ is about $20\%$, with a similar improvement for
the measurement of $\sin2\alpha$ in $\Bz\to\pip\pim$.  Belle also
measure $\calB(\Bz\to\etapr\piz) = (2.79 ^{+1.02 +0.25} _ {-0.96
  -0.34})\timesix$.

Decays like $\Bz\to\KS\KS\KS$ only proceed via a $b\to s\bar{s}s$
penguin diagram.  In these decays, SM pollution is therefore avoided
altogether, making them a very clean probe for new physics.
The related decay $\Bz\to\KS\KS\KL$ was studied by \babar.  It is
already experimentally known that the resonant contribution from
$\phi(\to\KS\KL)\KS$  to this decay is small, but the non-resonant component may be
large\cite{Cheng:2005ug}.  Assuming a uniform Dalitz distribution and
analysing $211 \invfb$, \babar\cite{Aubert:2006zy} sets a 90\% CL
upper limit of $\calB(\Bz \to\KS\KS\KL) < 6.4\timesix$.  Due to a low
product of efficiency and daughter branching fraction, this decay is
of limited use for the understanding of CP violation in $b\to\qqbar s$
decays. 

\section{Measurements related to $\alpha$}
Decays containing a $b\to u$ transition can be used to measure the
angle $\alpha \equiv \phi_2$ in the unitarity triangle. In general
several amplitudes contribute to these decays, only allowing the
direct measurement of an effective parameter $\alpha_\mathrm{eff}$.
There are several methods to extract the true angle $\alpha$ in
presence of this `pollution.'  For $\Bz\to\rhop\rhom$, isospin
symmetry in $B$ decays to $\rho\rho$ can be used to measure the shift
$2\delta\alpha$.  The previously available world averages for the
branching fractions\cite{hfag0705} were hard to reconcile with isospin
symmetry.  This has changed with new results from both $B$ factories:
the Belle collaboration\cite{Somov:2006sg} measures
$\calB(\Bz\to\rhop\rhom) = (22.8 \pm 3.8 ^{+2.3}_{-2.6})\timesix$.
\babar{} has a preliminary result of
$\calB(\Bp\to\rhop\rhoz) = (17.2 \pm 2.5 \pm 2.8)\timesix$.  Both
decays are found to be almost entirely longitudinally polarized.  With
the new branching fractions, the isospin triangles close.

Another new decay studied by \babar{} and Belle is $\Bz\to
a_1^\pm\pi^\mp$, from which $\alpha$ can be extracted up to a
four-fold ambiguity.  Exploiting isospin or approximate SU(3) flavor
symmetries this ambiguity can be overcome\cite{Gronau:2005kw}.  
This needs also the
measurement of related axial--vector decays, from which a
model-dependent measurement of $\alpha$ can be derived.  \babar{}
searches for $\Bz\to a_1^\pm\pi^\mp$ in $211\invfb$ and
measures\cite{Aubert:2006dd} a branching fraction of $\calB(\Bz\to
a_1^\pm\pi^\mp) = (33.2\pm3.8\pm3.0)\timesix$, assuming $\calB(a_1^+
\to (3\pi)^+) = 1$.  This is confirmed by Belle\cite{Abe:2005rf}. The
next step is to extend this analysis to measure time-dependent CP
violation in this decay.

\section{Other charmless $B$ decays}
The naive expectation for the longitudinal polarisation $f_L$ in $B$
decays into two vector mesons is $f_L \sim 1-m_V^2/m_B^2$, which is
fulfilled to a good approximation in tree-dominated decays such as $B
\to \rho \rho$.  There seems to be a pattern emerging where
$f_L$ is smaller than the naive expectation in decays dominated by loop
diagrams.  This was first seen in the decays
$B\to\phi\Kstar$ where $f_L\approx 0.5$.  To establish whether
loop-induced decays generally have a lower $f_L$, \babar{} has
searched\cite{Aubert:2006vt} for the related decays $B\to\omega V$,
where $V = \rho, \Kstar, \omega, \phi$.  Only $\Bp\to\omega\rhop$ was
observed with $\calB(\Bp\to\omega\rhop) = (10.6 \pm 2.1 ^{+1.6}
_{-1.0})\timesix$.  In this decay, $f_L = 0.82 \pm 0.11 \pm 0.02$ was
found.

In $B$ decays to final states comprising $\etaprp K^{(*)}$ the effect
of the $\eta$--$\etapr$ mixing angle combines with differing
interference in the penguin diagrams to suppress the final states
$\eta K$ and $\etapr \Kstar$, and enhance the final states $\etapr K$
and $\eta \Kstar$.  \babar{} finds evidence for the decays $B\to\etapr
\Kstar$ in 211\invfb and measures branching fractions of
$\calB(\Bp\to\etapr\Kstarp) = (4.9^{+1.9}_{-1.7}\pm0.8)\timesix$ and
$\calB(\Bz\to\etapr\Kstarz) = (3.8\pm1.1\pm0.5)\timesix$.  For the
related decays into $\etapr\rho$, only $\Bp\to\etapr\rhop$ is seen
with $\calB(\Bp\to\etapr\rhop) =
(6.8^{+3.2}_{-2.9}{}^{+3.9}_{-1.3})\timesix$, while $\Bz\to\etapr \rhoz$
is small with a 90\% C.L. upper limit of $\calB(\Bz\to\etapr\rhoz) <
3.7\timesix$.  Theoretical predictions using SU(3) flavor
symmetry\cite{Chiang:2003pm}, QCD factorization\cite{Beneke:2003zv},
and perturbative QCD factorisation\cite{Liu:2005mm} agree within
errors with the observed branching fractions.

\section{Summary}
Charmless hadronic $B$ decays provide a rich field for tests of QCD
and the standard model of electroweak interactions.  They allow to
constrain the SM contribution to $\Delta S_f$ in loop-dominated $B$
decays and precision tests of QCD models.  With the currently analyzed
statistics, decays with branching fractions of the order of $10^{-6}$
are within experimental reach.

\end{document}